\documentclass[letterpaper,10pt,conference]{ieeeconf}
\IEEEoverridecommandlockouts
\let\labelindent\relax

\usepackage{etex}
\usepackage[dvipsnames]{xcolor}
\usepackage{amsmath}
\usepackage{amsfonts}
\usepackage{amssymb}
\usepackage{amsthm}
\usepackage{mathrsfs}
\usepackage{siunitx}
\usepackage{verbatim}   
\usepackage{mathtools}	
\usepackage{siunitx}	
\usepackage{mdframed}
\usepackage{graphicx}

\usepackage[activate={true,nocompatibility},final,tracking=true,kerning=true,spacing=true,factor=1100,stretch=10,shrink=10]{microtype}
\usepackage[colorlinks,hidelinks]{hyperref}

\usepackage[style=ieee, backend=biber]{biblatex}

\usepackage{array}
\usepackage{multirow}
\usepackage{enumerate}
\usepackage{booktabs}

\usepackage{enumitem}
\usepackage{cuted}
\usepackage{arydshln}

\graphicspath{{./fig/}}

\usepackage{tikz}
\usetikzlibrary{arrows}
\usetikzlibrary{decorations.pathmorphing}
\usetikzlibrary{decorations.markings}
\usetikzlibrary{matrix}
\usetikzlibrary{calc}
\usetikzlibrary{patterns}
\usetikzlibrary{positioning}
\usetikzlibrary{shapes}
\usetikzlibrary{shapes.symbols}
\usetikzlibrary{shapes.geometric}
\usetikzlibrary{shadows.blur}
\usetikzlibrary{shapes.arrows}
\usetikzlibrary{shapes.multipart}
\usetikzlibrary{shapes.callouts}
\usetikzlibrary{shapes.misc}

\tikzstyle{every node}=[font=\small]
\tikzstyle{every path}=[line width=0.8pt,line cap=round,line join=round]

\DeclareMathAlphabet{\mathbbold}{U}{bbold}{m}{n}

\newcommand{\zero}{\mathbbold{0}}

\newcommand{\real}{\mathbb{R}}

\newcommand{\realnonnegative}{\mathbb{R}_{\geq0}}
												
\newcommand{\complex}{\mathbb{C}}

\newcommand{\setdef}[2]{\{#1 \;|\; #2\}}

\renewcommand{\j}{\boldsymbol{\mathrm{j}}}

\newcommand{\vect}[1]{\mathbbold{#1}}
\newcommand{\vones}[1][]{\vect{1}_{#1}}

\DeclareSymbolFont{bbold}{U}{bbold}{m}{n}
\DeclareSymbolFontAlphabet{\mathbbold}{bbold}

\newcommand{\map}[3]{#1: #2 \rightarrow #3}

\renewcommand{\theenumi}{(\roman{enumi})}

\newcommand{\T}{\mathsf{T}} 

\newcommand\oprocendsymbol{\hbox{$\square$}}
\newcommand\oprocend{\relax\ifmmode\else\unskip\hfill\fi\oprocendsymbol}

\newtheorem{theorem}{Theorem}[]
\newtheorem{lemma}{Lemma}[]

\newtheorem{proposition}{Proposition}

\theoremstyle{definition}
\newtheorem{definition}{Definition}[]

\newtheorem{problem}{Problem}[]

\theoremstyle{remark}

\newenvironment{pfof}[1]{\vspace{1ex}\noindent{\itshape Proof of #1:}\hspace{0.25em}} {\hfill\oprocend\vspace{1ex}}
\renewenvironment{proof} {\noindent{\itshape Proof:}\hspace{0.25em}} {\hfill\oprocend\vspace{1ex}}

\makeatletter
\newcommand{\vast}{\bBigg@{4}}
\newcommand{\Vast}{\bBigg@{5}}

\newcommand{\define}{\ensuremath{\triangleq}}

\title{Data-Driven Output Regulation using Single-Gain Tuning Regulators}

\author{Liangjie Chen and John W. Simpson-Porco
    \thanks{The authors are with the Department of Electrical and Computer Engineering, University of Toronto, 10 King's College Road,
    Toronto, ON, M5S 3G4, Canada. Email: \texttt{\footnotesize liangjie.chen@mail.utoronto.ca}, \texttt{\footnotesize jwsimpson@ece.utoronto.ca}.
    }
}

\begin{document}

\maketitle
\pagestyle{empty}
\thispagestyle{empty}

\begin{abstract}
 Current approaches to data-driven control are geared towards optimal performance, and often integrate aspects of machine learning and large-scale convex optimization, leading to complex implementations. In many applications, it may be preferable to sacrifice performance to obtain significantly simpler  controller designs. We focus here on the problem of output regulation for linear systems, and revisit the so-called tuning regulator of E. J. Davison as a minimal-order data-driven design for tracking and disturbance rejection. Our proposed modification of the tuning regulator relies only on samples of the open-loop plant frequency response for design, is tuned online by adjusting a single scalar gain, and comes with a guaranteed margin of stability; this provides a faithful extension of tuning procedures for SISO integral controllers to MIMO systems with mixed constant and harmonic disturbances. The results are illustrated via application to a four-tank water control process.
%
\end{abstract}


 
\section{Introduction}\label{sec:Introduction}


Many multivariable controller design methods require, as a starting point, knowledge of a reasonably accurate parametric system model. 
Presently however, motivated by high-complexity and/or large-scale control problems where building or fitting a parametric model is prohibitively expensive, interest in direct model-free or \emph{data-driven} multivariable controller design methods is steadily increasing. Established control problems such as the LQR problem \cite{FLL-DV-KGV:12,BR:19,CDP-PT:20a,SD-HM-NM-BR-ST:20,PLD-MR-MF-JZK:21} and MPC \cite{JC-JL-FD:19,JB-JK-MAM-FA:21} have recently been investigated from a learning-based or data-based perspective. This paper instead examines a data-driven design method for the classical output regulation problem \cite{JH:04,AI:17}.


Popular technical approaches for data-based control include reinforcement learning \cite{NM-AP-AR-ST:19}, deep learning \cite{PLD-MR-MF-JZK:21}, and behavioural systems theory \cite{JC-JL-FD:19,BR:19,JB-JK-MAM-FA:21,IM-FD:21b}. Many of these approaches are geared towards obtaining optimal performance, and either incorporate machine learning modules or require the solution of other large convex optimization problems parameterized by collected data. For both practical and theoretical reasons, it important to consider the possibility of trading off performance for increased simplicity in both design and implementation of data-driven controllers. Perhaps the simplest and most successful controller of all, the proportional-integral (PI) controller, may be tuned in a learning-based fashion via the Ziegler–Nichols procedure, and involves no optimization or machine learning. This suggests that revisiting traditional control paradigms from decades past will shed light on the complexity-performance trade-off for data-driven control.

Motivated precisely by the model-independence and online tuning success of PI control, E. J. Davison in 1976 introduced the \textit{multivariable tuning regulator}, a minimal-order controller which solves the error-feedback output regulation problem for stable multi-input multi-output (MIMO) linear time-invariant (LTI) systems \cite{EJD:76}\cite[Chp. 4]{EJD-AGA-DEM:20}; see also \cite{DEM-EJD:89, DED-EJD:11, WSWW-DED-EJD:13, ATJ-VA-VDSM-CX:19} for related work. The design asymptotically rejects any combination of polynomial and harmonic disturbances, and enjoys two remarkable features: (i) it is inherently data-driven, as the feedback gain matrices depend  \emph{only} on frequency response data, and (ii) it can be systematically tuned online with a guarantee of closed-loop stability. Recently, similar properties have been established for integral control of nonlinear systems \cite{CG-HL-ST:17,JWSP:20a,MG-DA-VA-LM:22,PL-GW-VN:20}, and have found use in online feedback-based optimization \cite{GB-DLM-MHDB-SB-JL-FD:21,GB-MV-JC-ED:21,JWSP:22d}.

Like the PI controller, and in contrast to most current data-driven control approaches, the tuning regulator favors simplicity over optimality. Unfortunately, the original tuning regulator suffers from two major design drawbacks; as these are technical in nature and require further background on the tuning regulator concept to explain, we defer further discussion on them to Section \ref{sec:Review}. Our objective here is to revisit the tuning regulator as a simple canonical data-driven design for MIMO output regulation, address several deficiencies in the original design methodology, and lay groundwork for further exploration of the complexity-performance trade-off curve in data-driven control.

\emph{Contributions:} We propose and analyze the \emph{single-gain tuning regulator} (SGTR), a simple data-driven output-regulating controller for stable LTI systems. The SGTR improves upon the original tuning regulator design in two ways. First, the SGTR design relies only on \emph{open loop} frequency response data from the plant (Theorem \ref{Thm:L2}), which can be determined via simple experiments \cite{EJD:76}; the original tuning regulator requires repeated reidentification during the tuning process. Second, the SGTR can be tuned online by adjusting a single scalar $\epsilon > 0$, while the original design of \cite{EJD:76} requires tuning of (in general) many scalar gains. In contrast with \cite{EJD:76}, our design comes with a stability certificate (Lemma \ref{Lem:EquivStab}) that the dominant closed-loop eigenvalues have a stability margin of $\mathcal{O}(\epsilon)$. In this sense, the design provides a true extension of classical data-driven SISO integral controller tuning procedures to the multivariable case with mixed constant and harmonic disturbances. We illustrate our design on a problem of disturbance rejection for the four-tank control process of \cite{KHJ:00}.





%
\emph{Notation:} For a matrix $A \in \complex^{m \times n}$, $\mathrm{conj}(A)$ denotes its element-wise complex conjugate, $A^*$ denotes its Hermitian transpose, $A^{\T}$ is its transpose without conjugation, $\mathrm{vec}(A)$ denotes its column-wise vectorization, and $A^{\dagger}$ denotes its Moore-Penrose pseudoinverse. If $A$ is square, then $\mathrm{eig}(A)$ denotes the set of all distinct eigenvalues of $A$. The symbol $\otimes$ is the Kronecker product. 
Given row vectors $x_1,\ldots,x_n$ of size $m$, $\mathrm{col}(x_1,\ldots,x_n)$ denotes the associated $n \times m$ stacked column matrix. 
Finally, we say $\map{F}{\real_{\geq 0}}{\real^{n\times m}}$ is $\mathcal{O}(\epsilon)$ as $\epsilon \to 0^+$ if $\lim_{\epsilon \to 0^{+}}\Vert F(\epsilon)\Vert / \epsilon < \infty$.

\section{Review: Linear Output Regulation}
\label{sec:Review}

\subsection{General Problem Formulation}
\label{sec:RSMP}
Consider the finite-dimensional causal LTI plant
\begin{equation}\label{eq:LTI}
\mathcal{P}: \quad \begin{aligned}
\dot{x} &= Ax + Bu + B_dd, \qquad x(0) \in \real^n\\
e &= Cx + Du + D_dd
\end{aligned}
\end{equation} 
with state $x(t) \in \real^n$ and control input $u(t) \in \real^m$. The output $e \in \real^r$ with $r \leq m$ is a set of measurable \emph{error} variables (e.g., tracking errors) to be regulated to zero. We assume throughout this work that $A$ is Hurwitz stable, but that the matrices $(A,B,B_{d},C,D,D_{d})$ are otherwise unknown, as is the order $n$ of the plant. The problem of regulating a stable but otherwise unknown system arises frequently in applications, and examples include large-scale power systems frequency control \cite{JWSP-NM:20m}, active noise cancellation \cite{SP-MB:10}, and chemical process control \cite{JP-HNK:80}.

The exogenous input signal $d \in \real^{n_d}$ models disturbances to be rejected and reference signals to be tracked, and is assumed to be  generated by the LTI \emph{exosystem}
\begin{equation}\label{eq:Exo}
\dot{w} = Sw\,, \qquad d = Ew, \qquad w(0) \in \real^{n_w},
\end{equation}
with state $w \in \real^{n_w}$. We assume here that $S \in \real^{n_w \times n_w}$ has only semisimple eigenvalues on the imaginary axis, and let
\begin{subequations}\label{eq:MinimalPoly}
\begin{align}
    \mu_{S}(s) &= s (s^2+\omega_1^2) \cdots(s^2+\omega_{\ell}^2) \label{eq:MinimalPoly-1}
    \\
    &= (s-\lambda_0) (s-\lambda_1) (s-\lambda_1^*) \cdots
    \label{eq:MinimalPoly-2} 
\end{align}
\end{subequations}
denote the minimal
polynomial of $S$ with order $q \define 2\ell + 1$ and where $0 < \omega_1 < \omega_2 < \cdots < \omega_{\ell}$. Note that $\mu_{S}(s)$ has one root at $\lambda_0 \define 0$, and for $k \in \{1,\ldots,\ell\}$, one complex conjugate pair of roots at $\lambda_k = \j\omega_k$ and $\lambda_k^* = -\j\omega_k$. We let
$
\hat{P}(s) = C(sI_n-A)^{-1}B + D
$
denote the $r \times m$ transfer matrix of \eqref{eq:LTI} from $u$ to $e$. For $k \in \{0, \ldots, \ell\}$, then $\hat{P}( \lambda_k)$ is the frequency response of the plant $\mathcal{P}$ on the $u \mapsto e$ channel evaluated at the $k$th exosystem eigenvalue. 

The problem of \emph{error-feedback output regulation} is that of designing a dynamic controller for \eqref{eq:LTI}, processing $e(t)$ and producing $u(t)$, such that the closed-loop system is internally exponentially stable when $d \equiv 0$, and such that $\lim_{t\to\infty}e(t) = 0$ for all initial conditions and for all exogenous input signals $d(t)$ generated by \eqref{eq:Exo}. Achieving regulation \emph{robustly} with respect to variations in the plant data requires a canonical two-piece construction of the controller, consisting of an error-processing subsystem (the \emph{internal model} or \emph{servocompensator}), and a stabilizing compensator. Our preferred construction of the servocompensator follows \cite{EJD:76}, and has the advantage that the states of the resulting servocompensator are easily interpreted in relation to the exosystem dynamics. We refer the reader to \cite[Chapter 4.4]{AI:17} for another common alternative construction of the servocompensator. 

Based on \eqref{eq:MinimalPoly}, define $\phi_0 \define 0$, $g_0 = 1$, with $\Phi_0 \define \phi_0 \otimes I_r = \zero_{r\times r}$, $G_0 \define g_0 \otimes I_r = I_r$. 
For $k \in \{1,\ldots,\ell\}$, similarly define
\begin{equation}\label{eq:Phi-G-def}
\phi_k \define \begin{bmatrix}
    0 & 1 \\ -\omega_k^2 & 0
\end{bmatrix}, \quad 
g_k \define \begin{bmatrix}
    0 \\ 1
\end{bmatrix}
\end{equation}
with $\Phi_k \define \phi_k \otimes I_r$ and $G_k \define g_k \otimes I_r$. Finally, we let
\[
\begin{aligned}
\phi &\define \mathrm{blkdiag}(\phi_0,\phi_1,\ldots,\phi_{\ell}) \in \real^{q\times q},\\
g &\define \mathrm{col}(g_0,g_1,\ldots,g_{\ell}) \in \real^{q},
\end{aligned}\qquad\begin{aligned}
\Phi &\define \phi \otimes I_r\\
G &\define g \otimes I_r.
\end{aligned}
\]
By construction, $\mathrm{eig}(\Phi) = \mathrm{eig}(S)$, and $(\Phi,G)$ is controllable.
The servocompensator (i.e., internal model) is
\begin{equation}\label{eq:internal-model}
\dot{\eta} = \Phi\eta + Ge, \qquad \eta(0) \in \real^{rq},
\end{equation}
which processes the error signal $e$. Consider now the cascaded system consisting of \eqref{eq:LTI} and \eqref{eq:internal-model}, with input $u$ and outputs $(e,\eta)$. The cascade is stabilizable and detectable \textemdash{} and hence, there exists a compensator stabilizing the cascaded system and solving the regulation problem \textemdash{} if and only if \cite{EJD-AG:75}
\begin{equation}\label{eq:NonResonance}
\mathrm{rank}\,\begin{bmatrix}
A-\lambda I_n & B\\
C & D
\end{bmatrix} = n+r, \quad \text{for all}\,\, \lambda \in \mathrm{eig}(S).
\end{equation}
Since $A$ is Hurwitz, by row operations \eqref{eq:NonResonance} is equivalent to
\begin{equation}\label{eq:FullRank}
\mathrm{rank}\, \hat{P}(\lambda) = r\qquad \text{for all}\,\,\lambda \in \mathrm{eig}(S).
\end{equation}
The ``non-resonance'' condition \eqref{eq:FullRank} stipulates that the transmission zeros of the plant $\mathcal{P}$ on the $u\mapsto e$ channel are disjoint from the poles of the servocompensator. 

\subsection{Davison's Tuning Regulator}
\label{sec:TuningRegulator} 
In \cite{EJD:76}, E.~J.~Davison posed an important special case of the design approach in Section \ref{sec:RSMP}, inspired by classical online tuning approaches for integral controllers. As motivation, consider the SISO integral controller
$\dot{\eta} = e$, $u = -\epsilon\eta$, 
where $\epsilon \in \real$ is the gain. For stable SISO LTI processes, the online tuning procedure is to select $\epsilon$ such that $\mathrm{sign}(\epsilon) = \mathrm{sign}(\hat{P}(0))$, and slowly increase the magnitude of $\epsilon$ from a small value until the desired tracking performance is achieved.\footnote{If satisfactory performance cannot be achieved, then the plant requires additional stabilizing pre-compensation before tuning of the integral loop.} 
This approach has three key characteristics:
\begin{enumerate}[label=(C\arabic*)]
\item \label{C1} only the DC gain of the \emph{open-loop plant} is required;
\item \label{C2} a stable closed-loop system can be systematically obtained through tuning of a \emph{single} scalar parameter, and the dominant pole of the closed-loop system has a negative real part which is of $\mathcal{O}(\epsilon)$ as $\epsilon \to 0^+$ \cite{JWSP:20a};
\item \label{C3} the control implementation is simple and practical.
\end{enumerate}

The so-called multivariable tuning regulator of \cite{EJD:76} was Davison's effort to mirror the characteristics \ref{C1}--\ref{C3} in the MIMO LTI case, and for more general refrence/disturbance signals generated by \eqref{eq:Exo}, with the following design procedure.
For the exogenous input signals $d \define \sum_{i=0}^\ell d_i$, let $d_i$ be a constant signal if $i=0$, and a harmonic signal with frequency $\omega_i$ otherwise. Then, we require an integral controller 
\[
    \mathcal{C}_0: \quad \dot{\eta}_0 = e,
    \quad u_0 = -\epsilon_0F_0\eta_0
\]
to reject $d_0$, and a resonant controller 
\[
    \mathcal{C}_k: \quad \dot{\eta}_k = \Phi_k\eta_k + G_k e, \quad
    u_k = -\epsilon_kF_k\eta_k
\]
to reject $d_k$ for $k \in \{1, \ldots, \ell\}$, where $(\Phi_k, G_k)$ are as defined in \eqref{eq:Phi-G-def} and $\epsilon_k$ are tuning parameters. The matrix gains $F_k$ are constructed as follows. 
\begin{lemma}[Lemma 3, \cite{EJD:76}]\label{lemma:EJD-constant}
     Suppose that $d = d_0$ and the DC gain satisfies $\,\mathrm{rank}\,\hat{P}(0) = r$. If $F_0 = \hat{P}(0)^{\dagger}$, then there exists an $\epsilon^{\star}$ such that for all $\epsilon_0 \in (0,\epsilon^{\star}]$, the closed-loop system with $\mathcal{P}$ and $\mathcal{C}_0$ is internally exponentially stable.
\end{lemma}
\begin{lemma}[Lemma 4, \cite{EJD:76}]\label{lemma:EJD-harmonic}
    Suppose that $d = d_k$ is harmonic with frequency $\omega_k$ and the frequency response satisfies $\,\mathrm{rank}\,P(\j \omega) = r$. Let $F_k \define \begin{bmatrix}
        F_k^1 & F_k^2
    \end{bmatrix}$, where $F_k^1 \define 2\omega_k \mathrm{Im}[\hat{P}(\j \omega_k)]^{\dagger}$ and $F_k^2 \define 2\mathrm{Re}[\hat{P}(\j \omega_k)]^{\dagger}$. Then, there exists an $\epsilon^{\star}$ such that for all $\epsilon_k \in (0,\epsilon^{\star}]$, the closed-loop system with $\mathcal{P}$ and $\mathcal{C}_k$ is internally exponentially stable.
\end{lemma}
Lemma \ref{lemma:EJD-constant} allows us to construct the controller $\mathcal{C}_0$ and tune $\epsilon_0$  so that the closed-loop system performance is satisfactory, while temporarily disregarding the effects of the harmonic exogenous signals $d_1, \ldots, d_\ell$. Similarly, 
Lemma \ref{lemma:EJD-harmonic} allows us to construct $\mathcal{C}_k$ and tune $\epsilon_k$ while temporarily disregarding the effects of the other harmonic exogenous signals $\{d_i\}_{i \neq k}$ and the constant $d_0$. 
For more general exogenous disturbances with  constant and $\ell$ harmonic components, the design process requires the sequential application of Lemma \ref{lemma:EJD-constant}, then Lemma \ref{lemma:EJD-harmonic} $\ell$ times. For $k \in \{1,\ldots,\ell\}$, constructing the gain matrix $F_k$ thus requires the frequency response data of the closed-loop system consisting $\mathcal{P}, \mathcal{C}_0, \ldots, \mathcal{C}_{k-1}$. Evidently, as $\ell$ increases, the implementation of Davison's regulator becomes more cumbersome, and we can conclude that it does not in fact possess the characteristics \ref{C1}--\ref{C3}. Moreover, while the design procedure produces a stable closed-loop system, no results have been reported regarding the margin of stability.


\section{The Single-Gain Tuning Regulator}
\label{sec:Analysis}

\subsection{Problem Statement}
\label{sec:ProbStatement}

Our objective is to remedy the tuning and commissioning issues present in the original tuning regulator proposal, resulting in a procedure more directly analogous to the SISO tuning of integral loops described in Section \ref{sec:TuningRegulator}. 
Thus, our new tuning procedure should (i) produce a direct mapping from (samples of) open-loop plant frequency response data to some fixed controller gains, and (ii) the number of online tuning parameters should be reduced to a single scalar $\epsilon > 0$. To this end, consider the \textit{single-gain tuning regulator} (SGTR)
\begin{equation}\label{eq:TuningReg}
\boxed{
\dot{\eta} = \Phi\eta + Ge, \qquad
u = - F(\epsilon)\eta,
}
\end{equation}
where $(\Phi,G)$ are as defined in Section \ref{sec:RSMP}. The feedback gain $\map{F}{\real_{\geq 0}}{\real^{m\times rq}}$ belongs to the class $\mathcal{F}$ of continuous mappings which are $\mathcal{O}(\epsilon)$ as $\epsilon \to 0^+$. In particular, note that $F$ need not be a linear function of $\epsilon$. 

\begin{figure}[ht!]
\begin{center}
\tikzstyle{block} = [draw, fill=white, rectangle, 
    minimum height=3em, minimum width=10em, blur shadow={shadow blur steps=5}]
    \tikzstyle{hold} = [draw, fill=white, rectangle, 
    minimum height=2em, minimum width=3em, blur shadow={shadow blur steps=5}]
\tikzstyle{sum} = [draw, fill=white, circle, node distance=1cm, blur shadow={shadow blur steps=8}]
\tikzstyle{input} = [coordinate]
\tikzstyle{output} = [coordinate]
\tikzstyle{pinstyle} = [pin edge={to-,thin,black}]
\begin{tikzpicture}[auto, scale = 0.5, node distance=2cm,>=latex', every node/.style={scale=0.9}]
    \node [hold] (system) {Plant};
    \node [hold, left of=system, node distance=3cm] (C0) {SGTR};
          
    \node [hold, name=exo, above of=system, node distance=1.5cm] {Exosystem};
    \draw [thick, -latex] (exo) -- node[name=w] {$d$} (system.north);

    \draw[thick,-latex] (C0) -- node[pos=0.5] {$u$} (system);
    \node [output, name=output_y, right of=system, node distance = 2cm] {};
    \draw [thick,-latex] (system) -- node[name=e] {$e$} (output_y);
	
	\coordinate [below of=system, node distance=1cm, name=tmp, xshift=-2cm] {};
	\coordinate [left of=C0, node distance=2cm, name=tmp0] {};
	
    \draw[thick,-latex] (e) |- (tmp) -| (C0);    

\end{tikzpicture}
\caption{The single-gain tuning regulator.}
\label{Fig:SGTR}
\vspace{-1.5em}
\end{center}
\end{figure}
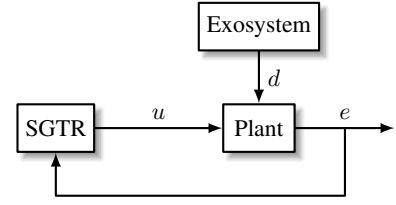

The architecture is shown in Figure \ref{Fig:SGTR}. Combining the SGTR \eqref{eq:TuningReg} with the plant $\mathcal{P}$ in \eqref{eq:LTI}, the closed-loop system takes the form
\begin{equation}\label{eq:OriginalClosedLoop}
\begin{aligned}
\begin{bmatrix}
\dot{x}\\
\dot{\eta}
\end{bmatrix} &= \underbrace{\begin{bmatrix}
A & -BF(\epsilon)\\
GC & \Phi - GDF(\epsilon)
\end{bmatrix}}_{\define \mathcal{A}(\epsilon)}\begin{bmatrix}
x\\
\eta
\end{bmatrix} + \begin{bmatrix}
B_w E \\ GD_w E
\end{bmatrix}w\\
e &= \begin{bmatrix}
C & -DF(\epsilon)
\end{bmatrix}
\begin{bmatrix}
x \\ \eta
\end{bmatrix} + \begin{bmatrix}
D_d E
\end{bmatrix}w
\end{aligned}
\end{equation}
with $w$ generated by \eqref{eq:Exo}. 
The presence of the servocompensator ensures that output regulation will be achieved if the closed-loop system is exponentially stable; we will omit the standard invariant subspace analysis \cite{AI:17}. The specific stability property we will seek to impose is inspired by the characteristic \ref{C2} of SISO integral control loops, as discussed in Section \ref{sec:TuningRegulator}. We let $\alpha(A) \define \max_{\lambda \in \mathrm{eig}
(A)} \mathrm{Re}[\lambda]$ denote the spectral abscissa of a square matrix $A$.


\begin{definition}[\bf Low-gain Hurwitz stability]\label{Def:LowGainStab}
A continuous matrix-valued function $\map{\mathcal{A}}{\real_{\geq 0}}{\real^{n \times n}}$ is \emph{low-gain Hurwitz stable} if there exist constants $c,\epsilon^* > 0$ such that $\alpha(\mathcal{A}(\epsilon)) \leq -c\epsilon$ for all $\epsilon \in [0,\epsilon^*)$. \hfill \oprocend
\end{definition}

Definition \ref{Def:LowGainStab} is stronger than the Hurwitz stability of $\mathcal{A}(\epsilon)$ for each $\epsilon \in (0,\epsilon^*)$, as the dominant eigenvalue of $A(\epsilon)$ is additionally required to be $\mathcal{O}(\epsilon)$ away from the imaginary axis for sufficiently small values of $\epsilon$. A Lyapunov characterization of low-gain Hurwitz stability is provided in Appendix \ref{sec:Proofs}.
We can now state our design problem.

\begin{problem}[\bf Single-gain tuning regulator]\label{Prob:SGTRD}
Given the minimal polynomial \eqref{eq:MinimalPoly} of the exosystem \eqref{eq:Exo} and the plant frequency response samples $\hat{P}(\lambda_k)$ for $k \in \{0,\ldots,\ell\}$, design a feedback $F \in \mathcal{F}$ such that the closed-loop system matrix in \eqref{eq:OriginalClosedLoop} is low-gain Hurwitz stable. \hfill \oprocend
\end{problem}


\subsection{Stability Analysis}
\label{sec:StabReduction}


%

%

We begin by developing a reduced characterization for low-gain Hurwitz stability of the closed-loop matrix $\mathcal{A}(\epsilon)$ from \eqref{eq:OriginalClosedLoop}. Given $A$ and $\Phi$, define the Sylvester operator
\begin{equation}\label{eq:SylTransform}
\map{\mathrm{Syl}_{\Phi,A}}{\real^{n\times rq}}{\real^{n\times rq}}, \; \mathrm{Syl}_{\Phi,A}(\Pi) \define \Pi \Phi - A\Pi.
\end{equation}
Since $\mathrm{eig}(\Phi)$ are imaginary and $A$ is Hurwitz, it is a standard result that $\mathrm{Syl}_{\Phi,A}$ is bijective \cite[Cor. A.1]{AI:17}, and we define an associated linear operator $\map{\mathscr{L}}{\real^{m\times rq}}{\real^{r \times rq}}$ as
\begin{equation}\label{eq:LoopOperator}
\begin{aligned}
&\mathscr{L}(F) = C\,\mathrm{Syl}_{\Phi,A}^{-1}(BF) + DF.
\end{aligned}
\end{equation}
Put differently, $\mathscr{L}(F) = C\Pi + DF$, where $\Pi \in \real^{n \times rq}$ is the unique solution to $\Pi\Phi - A\Pi = BF$. We call $\mathscr{L}$ the \emph{steady-state loop gain (SSLG)} operator of the system \eqref{eq:LTI} with respect to the exosystem \eqref{eq:Exo}. Our first key result is the following.

\begin{lemma}[\bf Reduction of SGTR stability analysis problem]\label{Lem:EquivStab}
The closed-loop system matrix $\mathcal{A}(\epsilon)$ in \eqref{eq:OriginalClosedLoop} is low-gain Hurwitz stable if 
\[
\mathcal{A}_{\rm red}(\epsilon) \define \Phi - G\mathscr{L}(F(\epsilon))
\] 
is low-gain Hurwitz stable.
\end{lemma}

\begin{proof}
Consider the Sylvester equation
\begin{equation}\label{eq:SylTransform2}
\mathrm{Syl}_{\Phi,A}(\Pi) = \Pi \Phi - A\Pi = - \tfrac{1}{\epsilon}BF(\epsilon),
\end{equation}
where $\Pi \in \real^{n \times rq}$, with unique solution
\[
\Pi(\epsilon) = \mathrm{Syl}_{\Phi,A}^{-1}(-\tfrac{1}{\epsilon}BF(\epsilon)) = -\tfrac{1}{\epsilon}\mathrm{Syl}_{\Phi,A}^{-1}(BF(\epsilon)),
\]
where in the second equality we have used linearity of $\mathrm{Syl}_{\Phi,A}$. Since $\epsilon \mapsto F(\epsilon)$ is continuous and is $\mathcal{O}(\epsilon)$ as $\epsilon \to 0^+$, we conclude that $\epsilon \mapsto \Pi(\epsilon)$ is continuous and is $\mathcal{O}(1)$ as $\epsilon \to 0^+$. Consider now the transformation matrix
\[
\mathscr{T} = \begin{bmatrix}
I_n & -\epsilon\Pi(\epsilon)\\
0 & \epsilon I_{rq}
\end{bmatrix}, \quad \mathscr{T}^{-1} = \begin{bmatrix}
I_n & \Pi(\epsilon)\\
0 & \tfrac{1}{\epsilon}I_{rq}
\end{bmatrix},
\]
which defines the change of state variables $(x^{\prime},\eta^{\prime}) = (x-\epsilon \Pi(\epsilon)\eta,\epsilon \eta)$. Direct computation shows that the system matrix $\mathcal{A}(\epsilon)$ from \eqref{eq:OriginalClosedLoop} transforms into
\[
\tilde{\mathcal{A}}(\epsilon) = \begin{bmatrix}
A - \epsilon\Pi(\epsilon) GC & \Pi(\epsilon) G C \mathscr{L}(F(\epsilon))\\
\epsilon GC & \mathcal{A}_{\rm red}(\epsilon).
\end{bmatrix}
\]
As $\mathcal{A}_{\rm red}(\epsilon)$ is low-gain Hurwitz stable, by Proposition \ref{Prop:LGHS} there exist constants $c_2^{\prime}, \epsilon^{\star} > 0$ and a continuous $P_2(\epsilon)$ which for all $\epsilon \in (0,\epsilon^{\star})$ satisfies $0 \prec P_2(\epsilon) \preceq c_2^{\prime} I_n$ and 
\[
\mathcal{A}_{\rm red}(\epsilon)^{\T}P_2(\epsilon) + P_2(\epsilon)\mathcal{A}_{\rm red}(\epsilon) = -\epsilon I_{rq}.
\]
Additionally, since $A$ is Hurwiz, there exists $P_1 \succ 0$ such that $A^{\T}P_1 + P_1A = -I_n$. Let $\tilde{\mathcal{P}}(\epsilon) = \mathrm{blkdiag}(P_1,P_2(\epsilon))$. Direct calculation then shows that for all $\epsilon \in (0,\epsilon^{\star})$,
\[
\begin{aligned}
\tilde{\mathcal{A}}(\epsilon)^{\T}\tilde{\mathcal{P}}(\epsilon) + \tilde{\mathcal{P}}(\epsilon)\tilde{\mathcal{A}}(\epsilon)
&= -\underbrace{\begin{bmatrix}
I_n + M_1(\epsilon) & -M_2(\epsilon)\\
 -M_2(\epsilon)^{\T} & \epsilon I_{rq}
\end{bmatrix}}_{\tilde{\mathcal{Q}}(\epsilon)},
\end{aligned}
\]
where
\[
\begin{aligned}
M_1(\epsilon) &= \epsilon P_1\Pi(\epsilon) GC + \epsilon(P_1\Pi(\epsilon) GC)^{\T}\\
M_2(\epsilon) &= P_1\Pi(\epsilon) G C \mathscr{L}(F(\epsilon)) + \epsilon (P_2 GC)^{\T}
\end{aligned}
\]
are both $\mathcal{O}(\epsilon)$ as $\epsilon \to 0^+$. It is clear that $\tilde{\mathcal{Q}}(\epsilon)$ is continuous, so again by Proposition \ref{Prop:LGHS}, it remains only to show that $\lambda_{\rm min}(\tilde{\mathcal{Q}}(\epsilon))$ is positive and $\mathcal{O}(\epsilon)$ as $\epsilon \to 0$.  A direct argument using Schur complements quickly establishes this, and hence $\tilde{\mathcal{A}}(\epsilon)$ is low-gain Hurwitz stable. 
\end{proof}

%
%

Lemma \ref{Lem:EquivStab} is effectively a time-scale separation result: for small $\epsilon$, the closed-loop eigenvalues decouple into two groups, the first being the eigenvalues of the open-loop plant, and the second being the eigenvalues of $\mathcal{A}_{\rm red}(\epsilon)$; see \cite[Chapter 2]{PVK-HKK-JO:99} for detailed discussion on this point. The result implies that we may focus our attention on low-gain stability of the matrix $\mathcal{A}_{\rm red}(\epsilon) = \Phi - G\mathscr{L}(F(\epsilon))$.\footnote{The converse result of Lemma \ref{Lem:EquivStab} in fact holds as well, but will be of no use for us here.} We next consider what properties the pair $(\Phi,G)$ should possess; see

\begin{definition}[\bf Low-gain stabilizability]\label{Def:LowGainStabilizability}
Let $\mathsf{A} \in \real^{n\times n}$ and $\mathsf{B} \in \real^{n \times m}$. The pair $(\mathsf{A},\mathsf{B})$ is \emph{low-gain stabilizable} if there exists a feedback $\mathsf{K} \in \mathcal{F}$ such that $\mathsf{A} - \mathsf{B}\mathsf{K}(\epsilon)$ is low-gain Hurwitz stable.
\end{definition}

This stabilizability property is characterized as follows; the proof can be found in the appendix.

\begin{lemma}[\bf Low-gain stabilizability]\label{Lem:Stabilizability}
A pair $(\mathsf{A},\mathsf{B})$ is low-gain stabilizable if and only if $(\mathsf{A},\mathsf{B})$ is stabilizable and all eigenvalues of $\mathsf{A}$ are contained in $\overline{\mathbb{C}}^-$.\footnote{This property is also known in the literature as \emph{asymptotic null-controllability with bounded controls} (ANCBC); see \cite{ZL:09}.}
\end{lemma}

By the constructions in Section \ref{sec:RSMP}, $(\phi,g)$ is controllable and all eigenvalues of $\phi$ are on the imaginary axis. We therefore conclude from Lemma \ref{Lem:Stabilizability} that $(\phi,g)$ is low-gain stabilizable. It follows that there always exists $z \in \mathcal{F}$ such that $\phi - gz(\epsilon)$ is low-gain Hurwitz stable, and with $Z(\epsilon) = z(\epsilon) \otimes I_r$, we immediately have that $\Phi - GZ(\epsilon)$ is low-gain Hurwitz stable. Comparing to $\mathcal{A}_{\rm red}(\epsilon)$ as defined in Lemma \ref{Lem:EquivStab}, we see that the question now becomes whether the linear operator equation $Z(\epsilon) = \mathscr{L}(F(\epsilon))$ can be solved for $F(\epsilon)$. If so, then we can first compute $Z(\epsilon)$, then recover a feedback gain $F(\epsilon)$ for use in \eqref{eq:TuningReg}. We summarize in a lemma, and move next to the study of the SSLG operator $\mathscr{L}$.

\begin{lemma}
Let $Z \in \mathcal{F}$ be such that $\Phi - GZ(\epsilon)$ is low-gain Hurwitz stable. If $Z(\epsilon) = \mathscr{L}(F(\epsilon))$ is solvable for $F \in \mathcal{F}$, then $F$ solves the SGTR problem.
\end{lemma}

\subsection{Computation of the Controller Gain \texorpdfstring{$F(\epsilon)$}{F(e)}}
\label{sec:SSLoopOperator}

Given $Z$, our goal is now to solve the operator equation $Z = \mathscr{L}(F)$ for $F$; indeed, this is always possible under \eqref{eq:FullRank}.

\begin{proposition}[\bf Surjectivity of SSLG operator]\label{Prop:L}
The SSLG operator $\mathscr{L}$ defined in \eqref{eq:LoopOperator} is surjective if and only if \eqref{eq:FullRank} holds. If in addition $r = m$, then $\mathscr{L}$ is invertible.
\end{proposition}

\begin{proof} 
The operator $\mathscr{L}$ is surjective if for any $Z \in \real^{r \times rq}$ there exists a solution $(\Pi,H)$ to
\begin{equation}\label{eq:BigL}
\begin{aligned}
\Pi \Phi &= A\Pi + BH\\
Z &= C\Pi + DH
\end{aligned}
\end{equation}
which we can equivalently write as
\[
\begin{bmatrix}0 \\ Z\end{bmatrix} = \begin{bmatrix}
A & B\\
C & D
\end{bmatrix}\begin{bmatrix}
\Pi \\ H
\end{bmatrix} + \begin{bmatrix}
I & 0\\
0 & 0
\end{bmatrix}\begin{bmatrix}
\Pi \\ H
\end{bmatrix}(-\Phi).
\]
This is a Hautus equation, and since $\mathrm{eig}(S) = \mathrm{eig}(\Phi)$, \cite[Thm. A.1]{AI:17} now yields that $\mathscr{L}$ is surjective if and only if \eqref{eq:NonResonance} holds, and hence if and only if \eqref{eq:NonResonance} holds. 
\end{proof}

Combining all results thus far, we can state the following.

\begin{theorem}[\bf Solvability of SGTR design problem]
Problem \ref{Prob:SGTRD} is solvable if \eqref{eq:FullRank} holds.
\end{theorem}

While the definition of $\mathscr{L}$ in \eqref{eq:LoopOperator} suggests that $\mathscr{L}$ depends on \emph{all the plant data} $(A,B,C,D)$, we will demonstrate that, in fact, $\mathscr{L}$ depends \emph{only} on the frequency response samples $\hat{P}(\lambda_k)$ and on the eigendecomposition of $\phi$; this enables \emph{gain computation based only on frequency response data}.

Recall from \eqref{eq:MinimalPoly} that $\{\lambda_0,\lambda_1,\lambda_1^*,\ldots,\lambda_{\ell},\lambda_{\ell}^*\}$ denote the roots of the minimal polynomial $\mu_{S} = \mu_{\Phi} = \mu_{\phi}$, and $q = 1 + 2 \ell$. 
Since the roots are all simple and distinct,  $\phi$ admits an eigen-decomposition $\phi = V\Lambda V^{-1}$ with eigenvalues $\Lambda \define \mathrm{diag}(\lambda_0, \lambda_1, \lambda_1^{\ast}, \ldots, \lambda_\ell, \lambda_\ell^{\ast})$ and right and left eigenvectors 
\[
\begin{aligned}
V &\define \begin{bmatrix}v_0 & v_1 & \mathrm{conj}(v_1) & \cdots & v_{\ell} & \mathrm{conj}(v_{\ell})\end{bmatrix}\\
V^{-1} = W &\define \mathrm{col}(w_0,w_1,\mathrm{conj}(w_1),\ldots,w_{\ell},\mathrm{conj}(w_\ell))
\end{aligned}
\]
with $\{v_k\}$ being column vectors and $\{w_k\}$ being row vectors. Finally, define the matrices
\begin{equation}\label{eq:XkYk}
X_k \define v_k w_k, \quad
\boldsymbol{X}_k \define X_k \otimes I_r, \quad k \in \{0,\ldots,\ell\},
\end{equation}
and we can state the key result.

\begin{theorem}[\bf Characterization of SSLG operator]\label{Thm:L2}
The SSLG operator $\mathscr{L}$ defined in \eqref{eq:LoopOperator} is equivalently given by
\begin{equation}\label{eq:LReduced}
\mathscr{L}(F) = \hat{P}(0)F\boldsymbol{X}_0 + 2\sum_{k=1}^{\ell}\nolimits \mathrm{Re}\{\hat{P}(\j \omega_k)F\boldsymbol{X}_k\}.
\end{equation}
\end{theorem}



\begin{proof}
To begin, recall the Sylvester operator defined in \eqref{eq:SylTransform}; we claim that
\begin{equation}\label{eq:SylIntegral}
\mathrm{Syl}_{\Phi,A}^{-1}(BH) = \int_{0}^{\infty}e^{A\tau}BH e^{-\Phi \tau}\,\mathrm{d} \tau.
\end{equation}
Since $A$ is Hurwitz and all eigenvalues of $\Phi$ have zero real part, all elements of $t \mapsto e^{At}$ decay exponentially, while all elements of $t \mapsto e^{-\Phi t}$ grow at most polynomially; it follows that all elements of $t \mapsto e^{At}BHe^{-\Phi t}$ tend to zero exponentially fast as $t \to \infty$, and hence the right-hand side of \eqref{eq:SylIntegral} is well-defined. Setting $\Pi = \mathrm{Syl}_{\Phi,A}^{-1}(BH)$ we verify that
\[
\begin{aligned}
\Pi\Phi - A\Pi &= \int_{0}^{\infty}e^{A\tau}BH e^{-\Phi \tau}\Phi - Ae^{A\tau}BH e^{-\Phi \tau}\Phi\,\mathrm{d} \tau\\
&= -\int_{0}^{\infty}\frac{\mathrm{d}}{\mathrm{d}\tau} (e^{A\tau}BHe^{-\Phi \tau})\,\mathrm{d}\tau = BH,
\end{aligned} 
\]
where we have again used that $A$ is Hurwitz. Since $\mathrm{Syl}_{\Phi,A}$ is bijective, \eqref{eq:SylIntegral} is indeed the unique solution of $\mathrm{Syl}_{\Phi,A}(\Pi) = BH$. Inserting \eqref{eq:SylIntegral} into \eqref{eq:LoopOperator}, we find that
\begin{equation}\label{eq:LHTimeDomain}
\mathscr{L}(H) = \int_{0^-}^{\infty}P(\tau)He^{-\Phi \tau}
\,\mathrm{d}\tau,
\end{equation}
where $P(t) \define Ce^{At}B \vones[\geq 0](t) + \delta(t)D$ is the causal impulse response matrix of the plant $\mathcal{P}$ from input $u$ to output $e$. The integral \eqref{eq:LHTimeDomain} can be evaluated via Laplace transform theory and contour integration. Define the matrix-valued signal $M(t) \define P(t)He^{-\Phi t}$. The signal $t \mapsto \int_{0^{-}}^{t}M(\tau)\,\mathrm{d}\tau$ has a Laplace transform $\frac{1}{s}\hat{M}(s)$ which is analytic in $\complex_{>0}$, and the signal has a well-defined limit as $t \to \infty$. Thus, by the final value theorem,
\begin{equation}\label{eq:LimitHatf}
\mathscr{L}(H) = \lim_{t\to\infty} \int_{0^-}^{t}M(\tau)\,\mathrm{d}\tau = \lim_{s\to 0^+}\hat{M}(s).
\end{equation}
Since $M(t)$ is the product of the two causal signals $P(t)$ and $He^{-\Phi t}\vones[\geq 0]$, it follows by convolution (e.g., \cite[Section 11-5]{WRL:61}) and taking the limit as $s \to 0^+$ that
\begin{equation}\label{eq:LIntegral}
\mathscr{L}(H) = \frac{-1}{2\pi\j}\int_{\sigma-\j\infty}^{\sigma+\j\infty}\underbrace{\hat{P}(\xi)H(\xi I_{rq}-\Phi)^{-1}}_{\define \Gamma(\xi)}\,\mathrm{d} \xi,
\end{equation}
where $\sigma \in \real$ is chosen such that the vertical line $\setdef{\sigma + \j\omega}{\omega \in \real}$ is contained within the region of convergence of the transform $\hat{P}$ of $P$, which is a superset of $\setdef{s \in \complex}{\mathrm{Re}(s) > \alpha(A)}$. 
Select $\sigma \in (\alpha(A),0)$, and consider the closed clockwise-oriented contour in $\complex$ consisting of the vertical line $\setdef{\sigma + \j\omega}{\omega \in \real}$ completed by an infinite semi-circle to the right of the vertical line.
As the contour encloses only the singularities of $(\xi I_{rq}-\Phi)^{-1}$, by Jordan's Lemma and the Residue Theorem we obtain
\begin{equation}\label{eq:LResidue}
\begin{aligned}
\mathscr{L}(H) &= \mathsf{Res}_{\lambda_0}\{\Gamma(\xi)\} \\ 
&\phantom{=} + \sum_{k=1}^{\ell} \left(\mathsf{Res}_{\lambda_k}\{\Gamma(\xi)\} +\mathsf{Res}_{\lambda_k^*}\{\Gamma(\xi)\}\right),
\end{aligned}
\end{equation}
where $\mathsf{Res}_{\lambda}\{\cdot\}$ evaluates the residue at $\xi = \lambda$. Note that
\[
\begin{aligned}
(\xi I_{rq} - \Phi)^{-1} &= V (\xi I_{q} - \Lambda)^{-1}V^{-1} \otimes I_r\\
&= \tfrac{1}{\xi-\lambda_0}\boldsymbol{X}_0 + \sum_{k=1}^{\ell}\left[\tfrac{\boldsymbol{X}_k}{\xi-\lambda_k} + \tfrac{\mathrm{conj}(\boldsymbol{X}_k)}{\xi-\lambda_k^*}\right].
\end{aligned}
\]
Since all poles of $\hat{P}(\xi)$ belong to $\mathbb{C}^-$ and all eigenvalues of $\Phi$ are simple, the residues evaluate to
\[
\begin{aligned}
\mathsf{Res}_{\lambda_0}\{\Gamma(\xi)\} &= \hat{P}(0)H\boldsymbol{X}_0\\
\mathsf{Res}_{\lambda_k}\{\Gamma(\xi)\} &= \hat{P}(\lambda_k)H\boldsymbol{X}_k\\
\mathsf{Res}_{\lambda_k^*}\{\Gamma(\xi)\} &= \mathrm{conj}(\hat{P}(\lambda_k)H\boldsymbol{X}_k),
\end{aligned}
\]
where we have used the fact that $\hat{P}(\lambda_k^*) = \mathrm{conj}(\hat{P}(\lambda_k))$. This leads immediately to \eqref{eq:LReduced} by combining terms.
\end{proof}

%
%
%
%


\subsection{SGTR Design Procedure}\label{sec:DesignProcedure}

The following three-step procedure provides a constructive solution to the design of the single-gain tuning regulator \eqref{eq:TuningReg}:
\begin{enumerate}\itemsep=4pt
\item[1)] Design $Z \in \mathcal{F}$ such that $\Phi - GZ(\epsilon)$ is low-gain Hurwitz stable; such a design always exists, since $(\Phi,G)$ is low-gain stabilizable (Lemma \ref{Lem:Stabilizability}). A particular approach which results in a low-dimensional design problem is to design $z \in \mathcal{F}$ such that $\phi - gz(\epsilon)$ is low-gain Hurwitz stable, and then simply set $Z(\epsilon) = z(\epsilon) \otimes I_r$. 
\item[2)] Solve the linear matrix equation $\mathscr{L}(F(\epsilon)) = Z(\epsilon)$ for $F \in \mathcal{F}$; a solution always exists since $\mathscr{L}$ is surjective (Proposition \ref{Prop:L}). The solution can be computed, for instance, by solving the vectorized linear system $\boldsymbol{M}\,\mathrm{vec}(F(\epsilon)) = \mathrm{vec}(Z(\epsilon))$, where
\[
\boldsymbol{M} = \boldsymbol{X}_0^{\T} \otimes \hat{P}(0) + 2\sum_{k=1}^{\ell}\nolimits \mathrm{Re}\{\boldsymbol{X}_k^{\T} \otimes \hat{P}(\j\omega_k)\}.
\]
\item[3)] Tune $\epsilon > 0$ for performance. By construction, there exists $\epsilon^{\star} > 0$ such that the closed-loop system will be internally exponentially stable for all $\epsilon \in (0,\epsilon^{\star})$.
\end{enumerate} 

As an example of what could be done in
step 1) above, one could pursue a pole-placement design by specifying that $\phi - gz(\epsilon)$ have a characteristic polynomial of the form
\begin{equation}\label{Eq:Desired}
(s+k_0\epsilon)(s+k_1\epsilon+\j\omega_1)(s + k_1\epsilon - \j\omega_1) \cdots
\end{equation}
for some positive constants $k_0, k_1$, and so on. This leads to an \emph{a priori} specified pattern of $\mathcal{O}(\epsilon)$ eigenvalues for the reduced system matrix $\mathcal{A}_{\rm red}(\epsilon)$ of Lemma \ref{Lem:EquivStab}. Explicit computation of feedback gains achieving desired pole placements, along with optimal designs, will be pursued in a future publication.

\section{Application: Four Tank Process}
\label{sec:Example}

To illustrate the ideas and to compare our single-gain regulator to Davison's original design, we consider a problem of disturbance rejection on the four-tank system of \cite{KHJ:00}, linearized at the operating point with minimum phase characteristics.
The control inputs $u(t) \in\real^2$ are the voltages applied to the two pumps, and the error output $e(t) \in \real^2$ is the deviation in tank water level measurements from their respective operating points. The exosystem is assumed to generate a constant disturbance and harmonic disturbances at $\omega_1 = 0.01$ rad/s and $\omega_2 = 0.1$ rad/s, together they model an external flow of water into tank 4. The minimum polynomial of $S$ therefore has the form
$\mu_{S}(s) = s(s^2+\omega_1^2)(s^2+\omega_2^2)$.

For the SGTR design, we follow the steps laid out in Section \ref{sec:DesignProcedure}. The intermediate feedback variable $z(\epsilon)$ is computed via pole placement such that $\mathrm{eig}(\phi - gz(\epsilon)) = \{-k_1\epsilon, -k_2\epsilon \pm \j \omega_1, -k_3\epsilon \pm \j\omega_2\}$. We then solve for $F(\epsilon)$ as described in the second step. Based on the trade-off between the overshoot and oscillatory behavior of the error trajectories, we select $\epsilon = 0.0002$, and $k_1 = 6.21, k_2 = 28.42, k_3 = 30.77$. 
 For Davison's design, we  follow the sequential procedure outlined in Section \ref{sec:TuningRegulator}, including recomputing of frequency response data after each loop is closed; we emphasize that the SGTR \emph{does not} require this extra burden. The tuned values obtained are $\epsilon_0 =\epsilon_1 = 0.0025$ and $\epsilon_2 = 0.003$.

\begin{figure}
    \centering
    \includegraphics[scale=0.85]{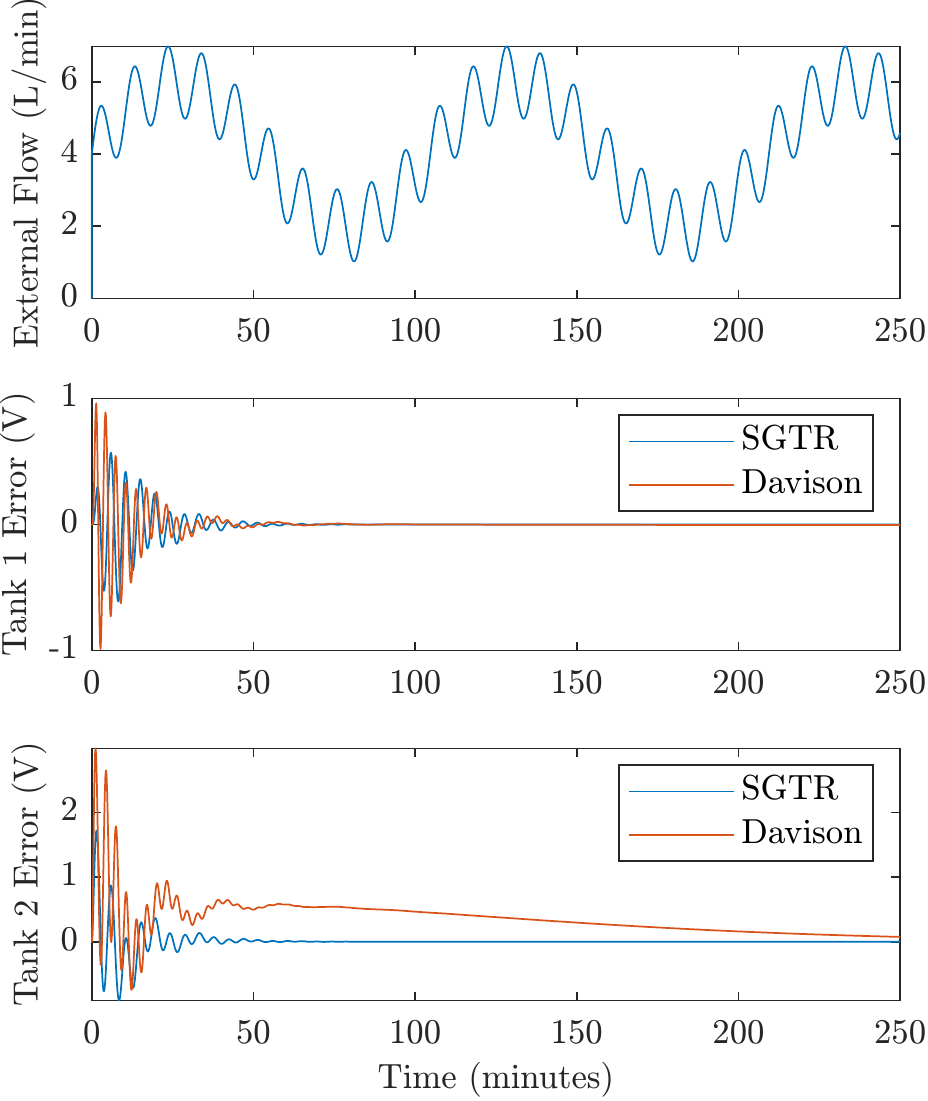}
    \vspace{-0.5em}
    \caption{Simulation result for the four-tank process.}
    \label{fig:sim_result}
\end{figure}

\begin{figure}
    \centering
    \includegraphics[scale=0.85]{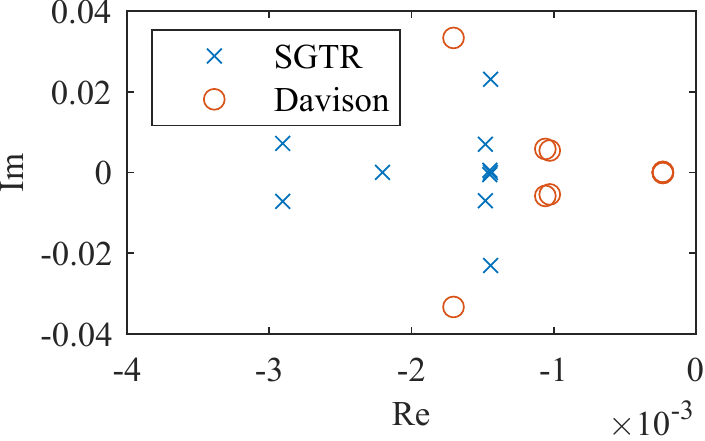}
    \vspace{-0.5em}
    \caption{Selected closed-loop system eigenvalues.}
    \label{fig:eig}
    \vspace{-1em}
\end{figure}

Figure \ref{fig:sim_result} shows the external flow disturbance $d(t)$ that enters the upper tank, and the closed-loop error trajectories in the two lower tanks. Our best tuning of Davison's design leads to a slower dominant mode, as can be seen in the error response for tank 2. The sequential tuning of $\{\epsilon_0,\epsilon_1,\epsilon_2\}$ in Davison's design leads to unnecessary performance trade-offs; for example, an increased value $\epsilon_0 = 0.005$ provides improved step disturbance rejection, but results in a smaller  range of stabilizing selections for $\epsilon_1$ and worse harmonic disturbance rejection. Figure \ref{fig:eig} shows\footnote{In practice, Figure \ref{fig:eig} would be impossible to produce due to the unknown plant dynamics, but is useful here for ground-truth comparison of the controllers.} the closed-loop system eigenvalues close to the imaginary axis for the two designs; the dominant eigenvalue with the SGTR is further to the left in $\complex^-$ than that with Davison's design.  

\section{Conclusions}
\label{sec:Conclusion}
We have proposed and developed a design procedure for the single-gain tuning regulator, which is a simple, data-driven, and minimal-order LTI controller solving the error-feedback output regulation problem for stable LTI systems. The design is based only on samples of the open-loop frequency response, is simple to compute, tune, and implement, and comes with a guaranteed stability margin. Several important directions for future work are being pursued, including extensions of the design procedure to the case of repeated exosystem poles and unknown exosystems \cite{AS-AI-LM:01},  incorporation of feedforward and proportional-derivative action, connections to more recent advances in data-driven control based on behavioral systems theory, discrete-time versions of the results, and applications in renewable energy integration problems.


\printbibliography

\appendices
\section{Supplementary Results and Proofs} \label{sec:Proofs}

Let $\mathbb{S}^{n}$ denote the set of $n \times n$ symmetric matrices. Let $\mathsf{Q}$ denote the set of maps $\map{Q}{\realnonnegative}{\mathbb{S}^{n}}$ with the property that there exist constants $\epsilon_{Q}^{\star}, c_{Q} > 0$ such that $Q$ and $Q(\epsilon) \succeq c_{Q}\epsilon I_n$ for all $\epsilon \in [0,\epsilon_{Q}^{\star}]$ and $Q$ is continuous on $[0,\epsilon_{Q}^{\star}]$. 
Let $\mathsf{P}$ denote the set of maps $\map{P}{\realnonnegative}{\mathbb{S}^{n}}$ with the property that there exists $\epsilon_{P}^{\star} > 0$ such that $P(\epsilon) \succ 0$ for all $\epsilon \in (0,\epsilon_{P}^{\star}]$ and $P$ is continuous on $[0,\epsilon_{P}^{\star}]$.

\begin{proposition}
[\bf Lyapunov for low-gain Hurwitz stability]
\label{Prop:LGHS}
A continuous matrix-valued mapping $\map{A}{\realnonnegative}{\real^{n \times n}}$ is low-gain Hurwitz stable if and only if for any $Q \in \mathsf{Q}$ there exists $P \in \mathsf{P}$ and $\epsilon_{\rm lyap}^{\star} > 0$  such that $A(\epsilon)^{\T}P(\epsilon) + P(\epsilon)A(\epsilon) = -Q(\epsilon)$ for all $\epsilon \in (0,\epsilon_{\rm lyap}^{\star})$.
\end{proposition}
The proof follows directly from the construction of $\mathsf{P}, \mathsf{Q}$ sets and textbook arguments on Lyapunov's direct method; we omit the details.  


\begin{pfof}{Lemma \ref{Lem:Stabilizability}}
(\emph{Necessity}) If $(\mathsf{A},\mathsf{B})$ is low-gain stabilizable, then there exists $\mathsf{K} \in \mathcal{F}$ and constants $c,\epsilon^{\star} > 0$ such that $\alpha(\mathsf{A}-\mathsf{B}\mathsf{K}(\epsilon)) \leq -c\epsilon$ for all $\epsilon \in [0,\epsilon^{\star})$. Hence, for any $\epsilon \in (0,\epsilon^*)$, $\mathsf{A} - \mathsf{B}\mathsf{K}(\epsilon)$ is Hurwitz, thus $(\mathsf{A},\mathsf{B})$ is stabilizable. Moreover, since $\mathsf{K}(\epsilon)$ is $\mathcal{O}(\epsilon)$ as $\epsilon \to 0^+$, we conclude that
$\lim_{\epsilon\to 0^+}\alpha(\mathsf{A}-\mathsf{B}K(\epsilon)) = \alpha(\mathsf{A}) \leq 0$.

\noindent (\emph{Sufficiency}) Since $\mathrm{eig}(\mathsf{A}) \subset \overline{\complex}^-$, partition the eigenvalues as $\mathrm{eig}(\mathsf{A}) \define \mathcal{E}_0 \cup \mathcal{E}_-$ where $\mathcal{E}_0$ are eigenvalues with zero real part and $\mathcal{E}_{-}$ are eigenvalues with negative real part. If $\mathcal{E}_0 \neq \emptyset$, then $\alpha(\mathsf{A}) = 0$, and $(\mathsf{A}, \mathsf{B})$ is stabilizable implies that all elements of $\mathcal{E}_0$ are controllable. We can then choose sufficiently small $c,\epsilon^{\star} > 0$ and use pole placement with a desired polynomial of the form \eqref{Eq:Desired} to construct a feedback $\mathsf{K}$ such that for all $\epsilon \in [0,\epsilon^{\star})$, $\alpha(\mathsf{A} - \mathsf{B}K(\epsilon)) \leq -c\epsilon$; continuity of $\mathsf{K}$ follows from continuity of the coefficients of \eqref{Eq:Desired} in $\epsilon$. If $\mathcal{E}_0 = \emptyset$, then we can choose sufficiently small $c,\epsilon^{\star} > 0$ and construct $\mathsf{K}$ such that $\alpha(\mathsf{A} - \mathsf{B}\mathsf{K}(\epsilon)) = \mathrm{Re}[\lambda_{\mathrm{dom}}]\leq -c\epsilon$ for all $\epsilon \in [0,\epsilon^{\star})$, where $\lambda_{\mathrm{dom}}$ denotes the dominant eigenvalue of $\mathsf{A}$, regardless of whether or not it is controllable. The continuity of $\mathsf{K}$ can be established similar to before.
%
\end{pfof}

\end{document}